Article

# Investigating the origin of topological-Hall-like resistivity in Zn-doped Mn$_2$Sb ferrimagnet

BoCheng Yu[1], JiaLiang Jiang[2], Jing Meng[1], XiaoYan Zhu[1,3], Jie Ma[4], HaiFeng Du[5,6,7], QingFeng Zhan[1], Jin Tang[2], Yang Xu[1]，and Tian Shang[1,8]

[1]Key Laboratory of Polar Materials and Devices (MOE), School of Physics and Electronic Science, East China Normal University, Shanghai, 200241, China

[2]School of Physics and Optoelectronic Engineering Science, Anhui University, Hefei, 230601, China

[3]State Key Laboratory of Infrared Science and Technology, Shanghai Institute of Technical Physics, Chinese Academy of Sciences, Shanghai, 200083, China

[4]Key Laboratory of Artificial Structures and Quantum Control, Shenyang National Laboratory for Materials Science, School of Physics and Astronomy, Shanghai Jiao Tong University, Shanghai, 200240, China

[5]Institutes of Physical Science and Information Technology, Anhui University, Hefei, 230601, China

[6]Anhui Province Key Laboratory of Condensed Matter Physics at Extreme Conditions, High Magnetic Field Laboratory, HFIPS, Anhui, Chinese Academy of Sciences, Hefei, 230031, China

[7]Science Island Branch of Graduate School, University of Science and Technology of China, Hefei, 230026, China

[8]Chongqing Key Laboratory of Precision Optics, Chongqing Institute of East China

Normal University, Chongqing, 401120, China

**Correspondence:** JiaLiang Jiang (jjl2024@ahu.edu.cn) | Yang Xu (yxu@phy.ecnu.edu.cn) | Tian Shang (tshang@phy.ecnu.edu.cn)

**Abstract**

Skyrmions and other chiral spin textures have been extensively studied as potential building blocks for novel spintronic devices. Hall-resistivity anomalies that deviate from magnetization scaling, known as the topological Hall effect, have been widely employed as evidence for the presence of chiral spin textures in magnetic materials. However, recent studies on magnetic thin films have revealed a drawback of this approach, as the presumed topological Hall contribution may in fact originate from trivial mechanisms. Here, we investigate the magnetic and transport properties of a Zn-doped $Mn_2Sb$ ferrimagnet, whose related compounds have previously been suggested to exhibit a topological Hall effect arising from chiral spin textures. Hall-resistivity anomalies are also observed in our sample, yet they show little correlation with the magnetic or metamagnetic transitions and are therefore clearly distinct from those in magnetic compounds hosting chiral spin textures. Most importantly, additional Lorentz transmission electron microscopy measurements rule out the existence of chiral spin textures in this ferrimagnet. Therefore, instead of a nontrivial origin, we attribute the Hall-resistivity anomalies to the combined effect of multiple anomalous Hall channels resulting from sample inhomogeneity. Our work shows that the difficulties of identifying chiral spin textures through transport measurements also apply to bulk systems, prompting some existing results to be revisited.

**1. Introduction**

Chiral spin textures (CSTs), exemplified by magnetic skyrmions, are soliton-like topological entities that may serve as novel information carriers. Owing to their small size, topological stability, tunability by external fields, etc. CSTs are very promising for the next-generation information-storage technologies [1-6]. Compared to the imaging or spectroscopic probes that are frequently used to identify the spin configurations in real space [7-12], magnetotransport measurements offer a more accessible approach to characterize the variation of electron wavefunctions perturbed by CSTs [13-18]. The conventional approach to detecting CSTs involves searching for additional contributions to Hall resistivity. In magnetic materials with topologically non-trivial spin textures, Hall resistivity is commonly expressed as $\rho_{xy} = \rho_{xy}^O + \rho_{xy}^A + \rho_{xy}^T$ [19-22], where the three terms represent the contributions from the ordinary (OHE), the anomalous (AHE), and the topological Hall effect (THE), respectively [19-27]. While the ordinary Hall term $\rho_{xy}^O$ ($= R_0 H$) is proportional to the applied magnetic field, the anomalous Hall term $\rho_{xy}^A$ ($= R_S M$) primarily scales with the magnetization, where $R_S$ is proportional to $\rho_{xx}^\alpha$ ($\alpha = 0, 1, 2$) [19-27]. Any additional contribution beyond these two terms, i.e., a nonzero $\rho_{xy}^T$, is often interpreted as evidence for the presence of CSTs, reflecting the Berry-phase accumulation of conducting electrons subject to the emergent magnetic field of the CSTs [19, 22, 28].

Recently, the validity of such an approach in revealing the CSTs has been questioned. It has been shown that the observed nonzero $\rho_{xy}^T$ may originate from trivial mechanisms rather than from topological ones, such as multiple $\rho_{xy}^A$ terms arising from sample inhomogeneity [29]. SrRuO$_3$ thin film and its related heterostructures are notable examples in which spatial variations of the electronic structure or magnetic properties have been found to account for the observed $\rho_{xy}^T$ [29 - 31]. Inhomogeneity is ubiquitous and also occurs in bulk single crystals, although its effects are generally less prominent than in thin-film systems. Therefore, extrinsic $\rho_{xy}^T$ is also an issue that calls for extra caution when searching for CSTs in bulk systems, yet it is much less explored.

The Mn$_2$Sb family of materials, where Mn atoms can be substituted by other $3d$ transition metals and Sb atoms can be substituted by Ga, Ge, As, Sn, and Bi [32-56], has been extensively studied for the magnetocaloric effect over the past decades [32, 37, 40, 44, 54, 55]. Very recently, this family of materials has regained research interest also because of the discovery of a nonzero $\rho_{xy}^T$ and related stripe magnetic domains, presumably due to CSTs [32, 46, 55]. Crystallizing in a centrosymmetric tetragonal structure (space group $P4/nmm$, $No.129$) [Fig.1**A**], the Mn$_2$Sb family of materials also provides a promising platform for investigating if CSTs can have trivial origins. CSTs have primarily been discussed in systems with Dzyaloshinskii-Moriya interactions, which arise from the breaking of inversion symmetry, either in the bulk crystals or at the interfaces of thin films [57-59]. However, some unconventional mechanisms, including the magnetic frustration and the competition between magnetic interactions and magnetic anisotropies, have been proposed to stabilize CSTs in certain centrosymmetric magnets [11, 60-67]. Compared to the noncentrosymmetric systems, CSTs in centrosymmetric systems exhibit unique advantages, e.g., tunable size and spin helicity [11]. Therefore, elucidating the nature of THE, as well as searching for CSTs in the centrosymmetric $Mn_2Sb$ family of materials, would be of considerable importance regarding both fundamental and technological aspects.

In this work, we performed both magnetization and electrical transport measurements on Zn-doped $Mn_2Sb$ single crystals. We observe THE-like anomalies in the Hall resistivity, consistent with previous results of other $Mn_2Sb$ family of materials [32, 46, 55]. However, the appearance of such anomalies showed little correlation with the magnetic or metamagnetic transitions across the studied temperature range. Subsequent Lorentz transmission electron microscope (LTEM) measurements exclude the existence of CSTs, indicating that the Hall-resistivity anomalies in the $Mn_2Sb$ family are most likely of extrinsic origin. Through aberration-corrected TEM (AC-TEM) and selected-area electron diffraction (SAED) characterization, we attribute the Hall-resistivity anomalies to the combined effect of multiple anomalous Hall loops induced by structural inhomogeneities.

## 2. Experimental details

The Zn-doped $Mn_2Sb$ single crystals were grown from Zn flux as reported in Ref. [32]. The typical dimensions of the obtained plate-like single crystals are approximately

$2\times2\times0.5$ mm$^3$ [see inset in Fig. 1**B**]. The crystal structure and phase purity were examined by powder x-ray diffraction (XRD) measurements, conducted on an Empyrean x-ray diffractometer with Cu K$\alpha$ radiation (Panalytical). Consistent with previous reports, our Zn-doped Mn$_2$Sb crystals are confirmed to crystallize in a centrosymmetric tetragonal structure. The XRD pattern obtained from a Mn$_{2-x}$Zn$_x$Sb single crystal exhibits a series of (00$l$) reflections [Fig. 1**B**], confirming its single-crystal nature and the orientation of the $c$-axis perpendicular to the crystal plane. The atomic ratios of the Mn$_{2-x}$Zn$_x$Sb single crystals were measured by energy dispersive x-ray spectroscopy (EDX), which confirms $x \approx 0.4$ and $0.55$ for the obtained crystals. Magnetization and electrical transport measurements were performed in a Quantum Design magnetic properties measurement system and physical property measurement system, respectively. For the resistivity measurements, electric current was applied in the $ab$-plane, while the magnetic field was applied along the $c$-axis.

Further characterizations involving imaging the magnetic domain configurations, transmission electron microscope (TEM) images and SAED patterns, were performed on a TEM instrument (Talos F200X, FEI) operated at an acceleration voltage of 200 kV. Fresnel magnetic images were recorded in Lorentz mode under an applied magnetic field along the $c$-axis and at temperatures down to 100 K. The thin lamellae with (001) planes, were fabricated via a standard lift-out method from the bulk using a focused ion beam instrument (Helios Nanolab 600i, FEI). The perpendicular magnetic field was adjusted by manipulating the objective current. A double-tilt liquid-nitrogen specimen holder (Model 636.6 cryotransfer holder, Gatan) was used to achieve temperatures down to 100 K. The high-angle annular dark-field scanning transmission electron microscopy (HAADF-STEM) images were obtained on an AC-TEM (Thermo Scientific Themis Z) equipped with a probe-forming spherical-aberration corrector.

## 3. Results and discussion

For the parent compound Mn$_2$Sb, the structure contains two crystallographically nonequivalent Mn sites [Mn(1) and Mn(2) in Fig. 1**A**], which enables various nearest- and next-nearest neighbor interactions, leading to rich magnetic properties. As shown in Fig. 1**C**, Mn$_2$Sb undergoes two magnetic transitions at $T_C \sim 550 - 600$ K and at $T_{SR} \sim 250$ K, which correspond to the paramagnetic (PM) to ferrimagnetic (FIM) and spin reorientation transitions, respectively [34]. Below $T_C$, magnetic moments in each of the Mn sublattices are ferromagnetically aligned along the $c$-axis but are antiparallel between Mn(1) and Mn(2) sublattices, leading to a FIM phase [denoted as FIM1, see Fig. S1**A** in the Supporting Information]. Below $T_{SR}$, magnetic moments in both Mn sublattices reorient in the $ab$-plane, thus forming another FIM phase [denoted as FIM2, see Fig. S1**B** in the Supporting Information]. The magnetic properties of Mn$_2$Sb can be effectively modulated by chemical substitutions on both Mn and Sb sites [32, 33, 35 - 56]. For example, in Mn$_{2-x}$Zn$_x$Sb, when the Zn content is increased, both $T_C$ and $T_{SR}$ are continuously suppressed, and finally, only a ferrimagnetic order with $T_C \sim$ 320 K persists for the $x = 1$ sample. For MnZnSb, the magnetic structure is similar to FIM1, except that the Mn(2) sublattice is completely replaced by nonmagnetic Zn atoms. As shown in Fig. 1**C**, our results are highly consistent with previous reports. Considering that Mn$_{2-x}$Zn$_x$Sb ($x \leq 0.8$) crystals exhibit similar magnetic properties, we performed most of the measurements on a Mn$_{1.6}$Zn$_{0.4}$Sb single crystal. Both the temperature-dependent magnetic susceptibility $\chi(T)$ and electrical resistivity $\rho_{xx}(T)$ reveal clear transitions at $T_C \sim 365$ K and $T_{SR} \sim 170$ K [indicated by dashed lines in Figs. 2**A**, **B**], respectively. The field-dependent magnetization $M(H)$ curves collected at various temperatures shown in Fig. 2**C** resemble typical features of ferrimagnet. As shown in Fig. 2**E**, the magnetic hysteresis becomes more evident as the temperature is lowered. Interestingly, in the FIM2 phase, as indicated by an arrow in Fig. 2**E**, the 10-K $M(H)$ exhibits a clear metamagnetic transition at $H_c \sim 0.23$ T. In contrast, in the FIM1 state [e.g., 200-K $M(H)$], this metamagnetic transition is absent. Figure 2**D** shows the field-dependent electrical resistivity $\rho_{xx}(H)$ at selected temperatures for Mn$_{1.6}$Zn$_{0.4}$Sb single crystal, here presented as magnetoresistivity (MR). Different from $M(H)$, no distinct anomalies can be identified in $\rho_{xx}(H)$. In the studied temperature range, the observed negative MR is consistent with the FIM nature of Mn$_{1.6}$Zn$_{0.4}$Sb. The MR at $\mu_0 H = 9$ T in the FIM2 phase is much larger than that in the FIM1 phase. As shown in Fig. 2**F**, the MR shows a sudden jump at $T > T_{SR}$. Since the Mn moments are oriented within the $ab$-plane in the FIM2 phase, the large negative MR reflects the suppression of magnetic scattering as the Mn moments align to the $c$-axis under an external field. While in the FIM1 phase, the Mn moments are parallel to the field direction, resulting in a relatively weak MR. The transition temperatures and magnetic fields determined from $\chi(T)$, $\rho_{xx}(T)$ and $M(H)$ are summarized in Fig. 3**A**.

To search for possible CSTs, we also performed field-dependent Hall resistivity $\rho_{xy}(H)$ measurements at various temperatures. As shown in Fig. 3**B**, the $\rho_{xy}(H)$ curves exhibit clear hysteresis loops, which are attributed to the FIM order in $Mn_{1.6}Zn_{0.4}Sb$. Consistent with the $M(H)$ data, the coercive field is nearly temperature-independent in the FIM1 phase, but it is significantly enhanced in the FIM2 phase. Interestingly, a peak-like feature was identified in the $\rho_{xy}(H)$ data, which becomes less pronounced but persists in the FIM2 phase (see Fig. S2 in the Supporting Information). Such Hall-resistivity anomalies have been previously observed in other members of the same material family, e.g., $Mn_{1.9}Co_{0.1}Sb$ [46], $Mn2Sb_{0.9}Bi_{0.1}$ [55], and $Mn_{1.15}Zn_{0.85}Sb$ [32], where CSTs were claimed as the origin for such anomalies. Here, the $\rho_{xy}(H)$ data can also be decomposed into different contributions. In Fig. 3**C**, we decompose the total Hall resistivity at 200 K into its individual components versus the magnetic field; a similar behavior is observed at other temperatures. The ordinary Hall effect can be described using either single-band or multi-band models. We applied both approaches and confirmed that the choice of model has a negligible influence on the magnitude and behavior of the extracted anomalous Hall signal. Moreover, we note that in Ref. [32], the single-band model was also employed to describe the ordinary Hall effect in materials with different compositions. Thus, for consistency and comparability, the results presented in this work are based on the single-band model. We use $\rho_{xx}^2$ to obtain $\rho_{xy}^A(H)$ from $\rho_{xy}^A(H) = \rho_{xx}^2 M(H)$ assuming a dominant intrinsic mechanism [19]. Considering that $\rho_{xy} \ll \rho_{xx}$ and the smooth evolution of $\rho_{xx}(H)$ curves, the use of $\rho_{xx}^2 M(H)$, $\rho_{xx} M(H)$, or simply $M(H)$ leads to comparable $\rho_{xy}^T$ values. After subtracting the $\rho_{xy}^O(H)$ and $\rho_{xy}^A(H)$ components from the measured $\rho_{xy}(H)$, the remaining contribution is shown as a dash-dotted line in Fig. 3**C**. The hump-like feature near zero field in $\rho_{xy}^T(H)$ resembles the THE claimed in other Mn$_2$Sb family of materials [32, 46, 55]. Thus, one may speculate that CSTs could also be present in Mn$_{1.6}$Zn$_{0.4}$Sb single crystals. According to our $\rho_{xy}^T(H,T)$ data, represented as a contour plot in Fig. 3**A**, $\rho_{xy}^T$ occurs over a broad range of temperatures and magnetic fields. Unlike the case in other skyrmion-lattice compounds [11, 60, 68 - 70], there is no apparent correlation between various magnetic phases and the appearance of $\rho_{xy}^T$, suggesting that the extracted $\rho_{xy}^T$ term likely has a trivial origin, rather than arising from topological CSTs, in Mn$_2$Sb family of materials [32, 46, 55].

To further understand the origin of the observed Hall resistivity anomalies, we performed high-resolution LTEM measurements on a Mn$_{1.6}$Zn$_{0.4}$Sb single crystal. The LTEM images collected at several representative temperatures across both the FIM1 and FIM2 phases in the absence of an external field are displayed in Figs. 4**A**-**C**. Upon applying a magnetic field, the LTEM contrast evolves similarly at all temperatures. Figures 4**E**-**H** show the field evolution in selected representative regions, using the 200 K data as an example, and additional field-dependent LTEM contrast data at different temperatures are provided in Figs. S3-S5 in the Supporting Information. The key observation from the LTEM measurements performed at various temperatures and magnetic fields, which cover the region of the magnetic phase diagram where a finite $\rho_{xy}^T$ was observed, is that no evidence of magnetic skyrmions or other CSTs could be discerned. Even in the field region where the extracted Hall anomaly is most pronounced (< 0.1 T), no magnetic domain changes or chiral spin textures were observed across the entire sample area in LTEM experiments (see Fig. S6 in the Supporting Information). These results unambiguously rule out the topological origin (i.e., spin chirality) for the extracted $\rho_{xy}^T$ in Mn$_{1.6}$Zn$_{0.4}$Sb.

Two types of stripe-like contrasts are observed in the LTEM images. The curved ones show no systematic evolution with applied field and are thus attributed to strain stripes. In contrast, the straight ones undergo a contrast reversal upon crossing a threshold field value, which varies across different sample regions. For example, in the region within the left white box in Fig. 4**B**, a bright stripe at 0.16 T (and lower fields) appears to shift to the left when the field is increased to 0.18 T (and remains basically unchanged at higher fields) [Figs. 4**E**, **F**]. In the region within the right white box, the bright stripe at 0.39 T appears to interchange with the adjacent dark stripe at 0.40 T [Figs. 4**G**, **H**]. The regional variation in the threshold field values indicates the presence of magnetic inhomogeneities in Mn$_{1.6}$Zn$_{0.4}$Sb. Further scanning transmission electron microscopy (STEM) measurements were performed to investigate such inhomogeneity. The straight stripes in the LTEM images are also present in the STEM images, suggesting that the magnetic contrast has a structural origin. Figure 4**I** shows the enlarged view of the area in Fig. 4**D** (marked by a white box). The serrated edges observed near the boundary between the stripe and the surrounding region suggest that the stripes may correspond to specific types of structural defects. Therefore, the stripes observed in the LTEM images are unlikely to be chiral domain walls, the latter representing another kind of CSTs giving rise to $\rho_{xy}^T$ [13-17, 71, 72].

To gain more insight into these stripes, we performed additional EDX measurements across a stripe, as indicated by a dashed arrow in Fig. 4I. The determined atomic fractions for all elements show negligible variation along the scan direction, suggesting a homogeneous composition across the stripes [Fig. 4J]. It is then natural to ask about the origin of such stripes in both the LTEM and STEM images. One possible scenario is that the stripes exhibit a crystal orientation distinct from that of the majority regions in the images. This implies that the orientations of the magnetization easy axis ($T > T_{SR}$) or easy plane ($T < T_{SR}$) relative to the direction of the applied magnetic field are also different from those of the majority regions. LTEM visualizes magnetic domains by detecting small deflections of the electron beam caused by the Lorentz force, which acts perpendicular to both the electron velocity and the local magnetization. If different regions exhibit distinct in-plane magnetization directions, the resulting electron deflection directions also differ. This variation in deflection leads to inhomogeneous electron distribution and produces contrast, thereby revealing different magnetic domains [11, 12, 73-75]. The inversion of black-white contrast observed in the LTEM images in Figs. 4E-H signifies a reversal of the net in-plane magnetic moment direction in these regions, thus supporting the hypothesis that the crystal orientation in the defect regions differs from that in the majority region. In simple terms, in the defect regions, the presence of grain boundaries between the defect and the majority region introduces additional stress, causing the magnetic moments to initially become pinned at a random angle—rather than aligning along the easy axis (for $T > T_{SR}$) or within the easy plane (for $T < T_{SR}$) as in the majority region. When a magnetic field is applied, it forces the pinned moments to revert to the direction of the easy axis or easy plane of the defect region. If the net in-plane magnetic moment before and after this reorientation points in opposite directions, a reversal of black-white contrast occurs. In contrast, the magnetic moment direction in the majority region remains unchanged with the applied field, and thus no contrast variation is observed. The details can also be found in Fig. S7 and Note S1 in the Supporting Information.

Measurements of atomic-resolution AC-TEM and SAED provide experimental evidence for the crystal orientation variations across different regions. Figure 5A presents a representative HAADF-STEM image containing both the majority regions and the stripes. In contrast to the well-defined atomic arrangement in the majority regions, the stripes exhibit a more complex atomic configuration. Figures 5B-D display magnified views of the left majority regions, stripes, and right majority regions, respectively. In the majority regions, atoms are arranged in a square lattice pattern, indicating that the [001] crystal orientation is parallel to the electron beam incidence. Conversely, atomic resolution is significantly reduced in the striated region due to the deviation of its [001] zone axis from the beam direction, resulting in overlapping projections of multiple atomic layers. This conclusion is further supported by the excellent agreement achieved when matching the observed HAADF-STEM images with simulated projections of the crystal structure rotated by a specific angle, unambiguously demonstrating the distinct crystal orientation of the stripes relative to the majority regions. Figure 5E displays a representative TEM image of an area comprising the majority regions and the stripes. Figures 5F-H present the corresponding SAED patterns acquired from the left majority region, stripes, and right majority region, respectively, as marked by the white rectangles in Fig. 5**E**. In contrast to the sharp diffraction spots observed in the majority region, the stripes exhibit elongated "comet-like" diffraction spots with pronounced tailing. When the electron beam irradiates multiple crystalline domains with slightly different orientations, each domain generates diffraction spots with minor positional offsets. If the misorientation angles are sufficiently small, these spots may partially overlap, resulting in elongated diffraction spots. This observation further confirms the distinct crystal orientation of the stripes relative to the majority region, unequivocally demonstrating the structural inhomogeneity of the material.

Crucially, as schematically illustrated in Fig. 6, this scenario also provides a natural explanation for the emergence of a finite $\rho_{xy}^T$. Due to the inhomogeneity of the local magnetic easy axis (plane), the defect and majority regions experience different effective local fields, despite the uniformity of the applied field. Locally, this alters the balance among several energy scales, namely the Zeeman energy associated with the alignment of each sublattice moment, the exchange energy between the sublattice moments, and the magnetic anisotropy energy. Overall, this results in inhomogeneous magnitudes, coercive fields, and polarities of the AHE loops contributed locally by the defect and majority regions. The combined effect of these AHE loops may manifest as a loop with low-field peaks, similar to the $\rho_{xy}^T$ term extracted in Fig. 3**C**. It is noted that, in practice, we cannot resolve the specific crystal orientation of each individual defect region. Therefore, it is not feasible to analyze this loop with low-field peaks using particular crystal orientations; instead, we propose one of the simplest possible combinations to explain the emergence of this loop. In this sense, the presence of finite $\rho_{xy}^T$ in Mn$_{1.6}$Zn$_{0.4}$Sb is unrelated to CSTs and instead has a trivial origin.

Although the present study primarily focuses on the Zn-doped Mn$_2$Sb sample with $x = 0.4$, our findings provide important insights applicable across a range of doping levels within this material system. As summarized in Supplementary Fig. S8, the magnetic properties of samples with different Zn concentrations ($x$ = 0.2 [48], 0.4, 0.55, 0.77 [32], and 0.85 [32]) evolve systematically with doping, whereas their Hall resistivity behavior remains qualitatively similar. Furthermore, direct comparison of the extracted anomalous Hall component and its temperature-field distribution for $x = 0.4$ and $x = 0.55$ (Supplementary Fig. S9) reveals nearly identical characteristics–both exhibiting a broad distribution of the additional Hall signal that is uncorrelated with specific magnetic phase boundaries. Given that the flux-growth method used for all these compositions inherently produces similar microstructural inhomogeneities, we reasonably conclude that the real-space magnetic imaging results would likewise be consistent across different doping levels. Therefore, we believe that the extrinsic origin of the Hall anomaly–rooted in sample inhomogeneity–represents a universal mechanism throughout the Zn-doped Mn$_2$Sb system.

Our conclusion regarding the extrinsic origin of the topological-Hall-like signal is specifically contextualized within the system with non-magnetic Zn doping. It is instructive to compare this case with systems involving magnetic dopants. For instance, in Co-doped Mn$_{1.9}$Co$_{0.1}$Sb, a large topological Hall signal has been reported and attributed to an intrinsic noncoplanar or noncollinear magnetic structure, arising from the competition of multiple magnetic states during spin reorientation [46]. Magnetic atomic doping like Co or Ni can introduce strong competing exchange interactions, significantly complicating the magnetic phase diagram and potentially stabilizing intrinsic topological spin textures. However, due to similarities in growth methods or sample characteristics, the observed Hall anomalies in such systems may also partially originate from extrinsic mechanisms such as sample inhomogeneity, making it challenging to disentangle intrinsic and extrinsic contributions. In contrast, non-magnetic Zn doping primarily alters the magnetic balance without introducing such additional magnetic interactions, resulting in a comparatively simpler magnetic structure. This provides a clearer platform to isolate and identify the extrinsic contribution. Therefore, the interpretation of topological Hall signals in doped magnetic materials must carefully consider the nature of the dopant.

## 4. Conclusion

In summary, we observed Hall resistivity anomalies in $Mn_{1.6}Zn_{0.4}Sb$ single crystals that extend beyond magnetization scaling. These anomalies have been speculated to be related to a topological Hall term associated with chiral spin textures formed around metamagnetic transitions [32, 46, 55]. However, we found that the Hall anomalies are not correlated with the magnetic phases. This finding rules out the existence of chiral spin textures and suggests that the Hall anomalies likely originate from the mimicry of the topological Hall effect caused by multiple anomalous Hall loops resulting from inhomogeneity. This is reminiscent of the well-known $SrRuO_3$ thin-film case, where inhomogeneity stems from variations in thickness and strain relaxation [29]. These sources of inhomogeneity, which are unique to thin films, are ineffective in bulk single crystals. By using a combination of real-space imaging techniques that are sensitive to local structural and magnetic properties, we identified structural defects with different crystal orientations and, therefore, different magnetic easy axes (planes), as the source of inhomogeneity in $Mn_{1.6}Zn_{0.4}Sb$. In addition, although a correlation between the transport and magnetic properties has been established in some other members of the $Mn_2Sb$ family of compounds, the similarity of their field-dependent Hall resistivity to our findings suggests that inhomogeneity may also affect those systems. Our results suggest that caution is advised when inferring the presence of chiral spin textures from transport measurements, not only in thin films but also in bulk materials.

**Author Contributions**

BochengYu: Single-crystal growth, XRD, resistivity measurements, data analysis, and writing–original draft. Jialiang Jiang: TEM, LTEM, AC-TEM, and EDX measurements. Jing Meng: Magnetization measurements. Xiaoyan Zhu: Magnetizaton and resistivity measurements. Jie Ma: Resistivity measurements. Haifeng Du: TEM, LTEM, AC-TEM, and EDX measurements. Qingfeng Zhan: Funding acquisition and project administration. Jin Tang: Funding acquisition and project administration. Yang Xu: Funding acquisition, project administration, and writing–review & editing. Tian Shang: Data analysis, funding acquisition, project administration, and writing–review & editing.

**Acknowledgements**

This work was supported by the National Key R&D Program of China (Grant No. 2022YFA1403603), the Natural Science Foundation of China (Grant Nos. 12374105, 12274125, 12574117, 12174103, 12174396, 12422403, U2032213, 12334008, 12350710785, and 12561160109), the Natural Science Foundation of Shanghai (Grant Nos. 21ZR1420500 and 21JC1402300), and the Natural Science Foundation of Chongqing (Grant No. CSTB-2022NSCQ-MSX1678). We thank Toni Shiroka for his insightful comments and suggestions.

**Conflict of Interests Statement**

The authors declare that they have no conflict of interest.

**Data Availability Statement**

The data that support the findings of this study are available from the corresponding author upon reasonable request.

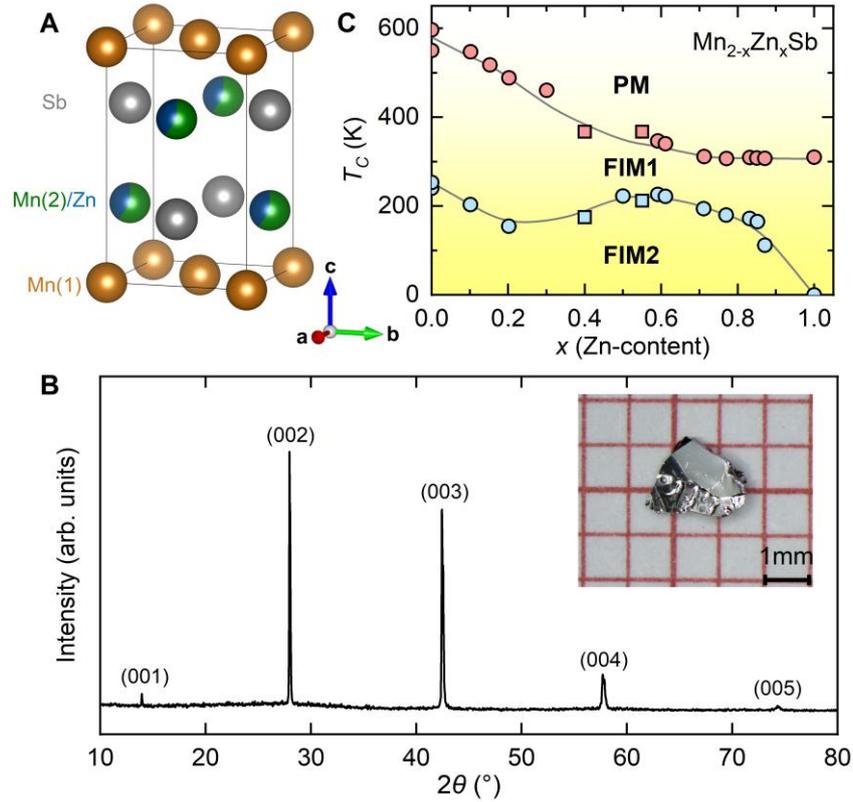

**Figure 1: The basic characterization of Mn$_{2-x}$Zn$_x$Sb.** **(A)** The crystal structure of Mn$_{2-x}$Zn$_x$Sb, demonstrating the substitution of Zn atoms for the Mn(II) atoms. **(B)** The XRD patterns collected on Mn$_{2-x}$Zn$_x$Sb single crystals, confirming their $(00l)$ orientation. The inset depicts a plate-like Mn$_{2-x}$Zn$_x$Sb ($x = 0.4$) single crystal. **(C)** Magnetic transition temperatures of Mn$_{2-x}$Zn$_x$Sb single crystals as a function of Zn-content $x$. The circlar symbols represent the data taken from Ref. [32-34], while cubic symbols represent data from this work. The PM, FIM1, and FIM2 denote the paramagnetic, and two different ferrimagnetic states, respectively.

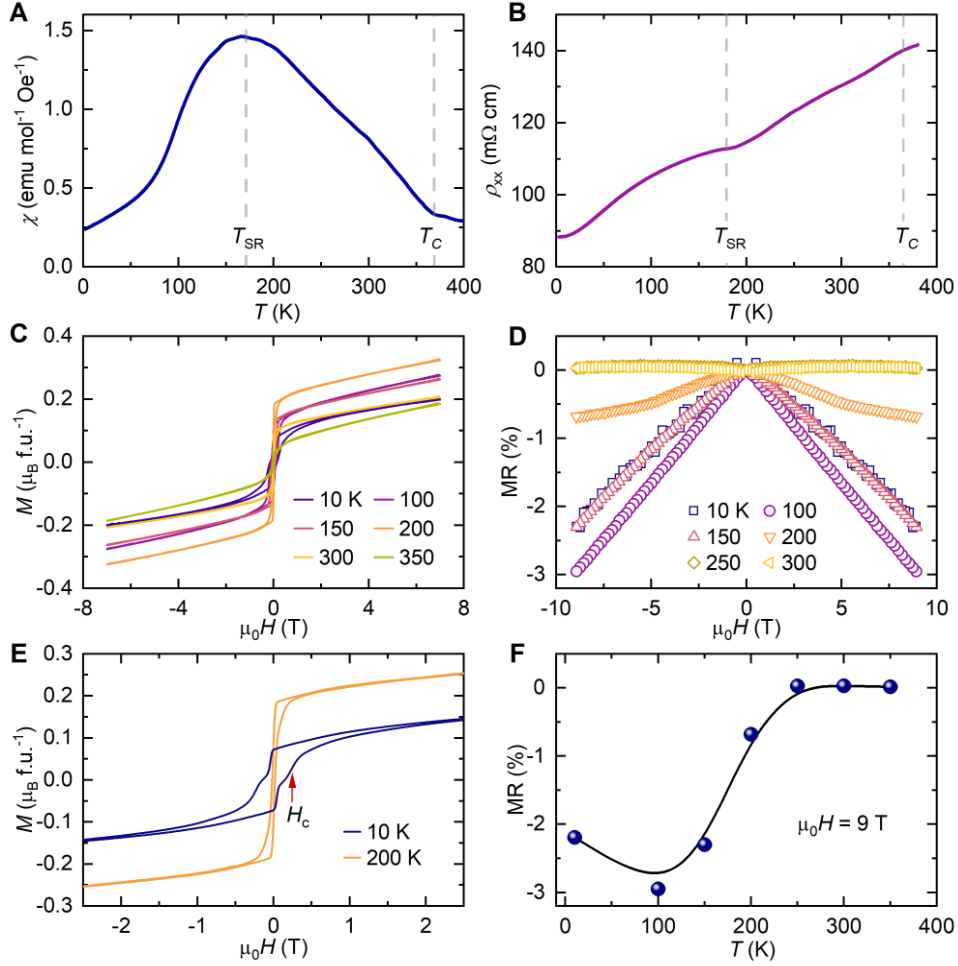

**Figure 2: Magnetic and transport properties of Mn$_{1.6}$Zn$_{0.4}$Sb. (A)** Temperature-dependent magnetic susceptibility $\chi(T)$ and **(B)** electrical resistivity $\rho_{xx}(T)$ of Mn$_{1.6}$Zn$_{0.4}$Sb single crystal. The $\chi(T)$ was measured at $\mu_0 H = 0.1$ T after zero-field cooling, while $\rho_{xx}(T)$ was measured in a zero-field condition. The dashed lines mark the FIM transition at $T_C$ and the spin reorientation transition at $T_{SR}$. **(C)** Field-dependent magnetization $M(H)$ and **(D)** electrical resistivity $\rho_{xx}(H)$ collected at various temperatures. Here, the $\rho_{xx}(H)$ is presented as magnetoresistivity, MR $= [\rho_{xx}(H) - \rho_{xx}(0)] / \rho_{xx}(0)$, where $\rho_{xx}(0)$ is the zero-field electrical resistivity. **(E)** Enlarged view of $M(H)$ at $T = 10$ K (in the FIM2 phase) and 200 K (in the FIM1 phase) in the low-field regime. The arrow marks the metamagnetic transition field $H_c$.

**(F)** The MR at $\mu_0 H = 9$ T as a function of temperature. Note that both magnetization and transport measurements were performed on the same piece of Mn$_{1.6}$Zn$_{0.4}$Sb single crystal.

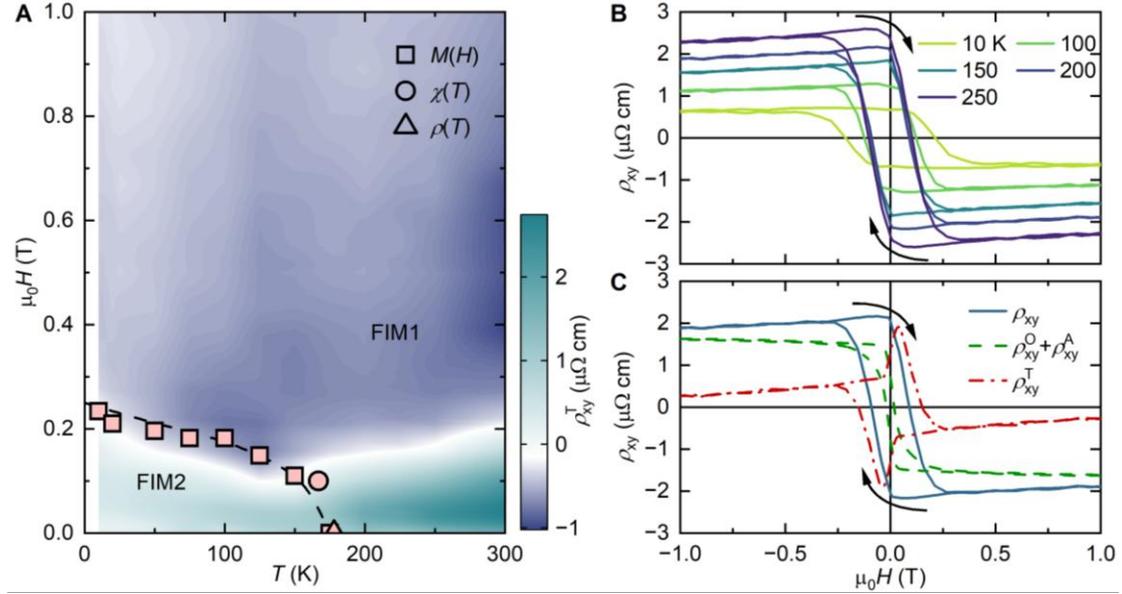

**Figure 3: Hall resistivity analysis and magnetic phase diagrams of Mn$_{1.6}$Zn$_{0.4}$Sb.**

**(A)** Magnetic phase diagram of a Mn$_{1.6}$Zn$_{0.4}$Sb single crystal with the field applied along the $c$-axis. The critical fields (square symbols) were determined from $M(H)$ data. The magnetic transition temperatures (triangle and circle symbols) were determined from the $\rho_{xx}(T)$ and $\chi(T)$ data. **(B)** The field-dependent Hall resistivity $\rho_{xy}(H)$ collected at various temperatures for Mn$_{1.6}$Zn$_{0.4}$Sb single crystal. **(C)** Analysis of the Hall resistivity at $T = 200$ K. Solid, dashed, and dash-dotted lines represent the measured total Hall resistivity $\rho_{xy}$, the sum of ordinary Hall resistivity and anomalous Hall resistivity ($\rho_{xy}^O + \rho_{xy}^A$), and the topological-Hall-like resistivity $\rho_{xy}^T$, respectively. The background color in panel **(A)** represents the magnitude of $\rho_{xy}^T$ at various temperatures.

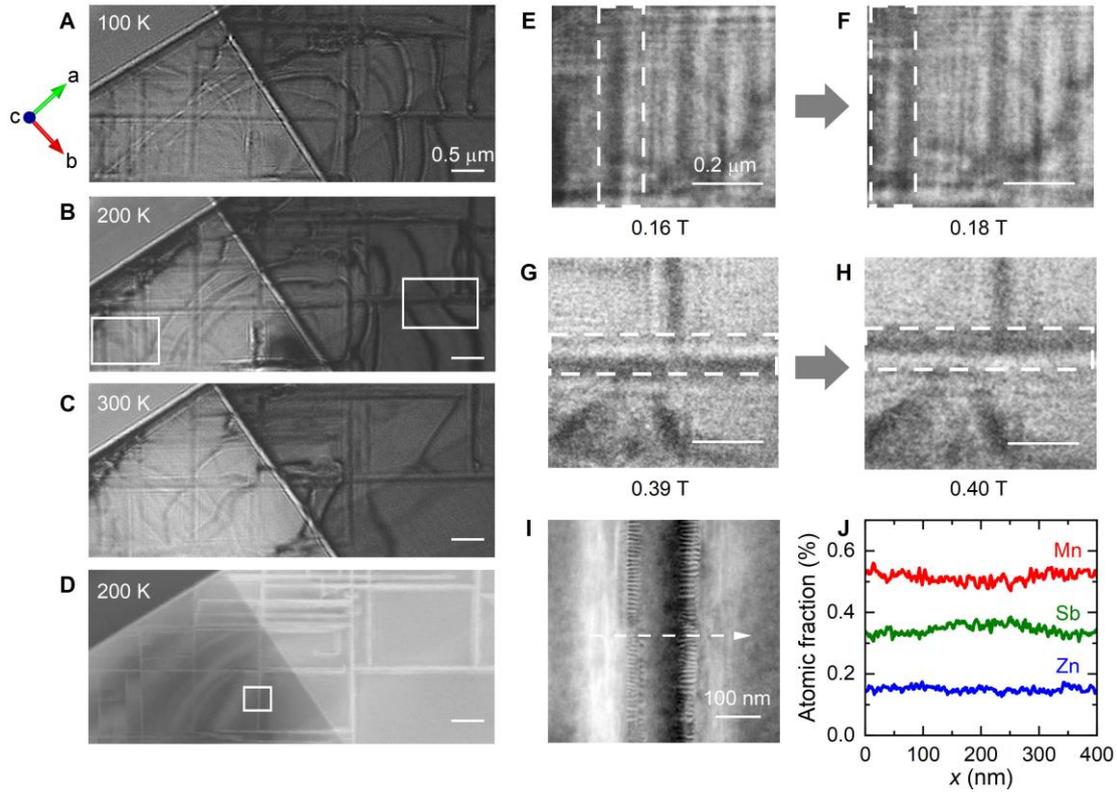

**Figure 4: Field-dependent stripe contrast reversal of Mn$_{1.6}$Zn$_{0.4}$Sb.** LTEM images of Mn$_{1.6}$Zn$_{0.4}$Sb collected at $T = 100$ K **(A)**, 200 K **(B)**, and 300 K **(C)** in the absence of an external field. The field evolution of the stripe contrast for the regions within the left and right white boxes in panel **(B)** is shown in panels **(E), (F)** and **(G), (H)**, respectively. The dashed boxes in panels **(E)-(H)** highlight the field-induced contrast changes. **(D)** STEM image collected at 200 K. **(I)** Enlarged view of the region marked by the white box in panel **(D)**. **(J)** Elemental distribution determined by EDX spectroscopy along the line scan indicated by the dashed arrow in panel **(I)**.

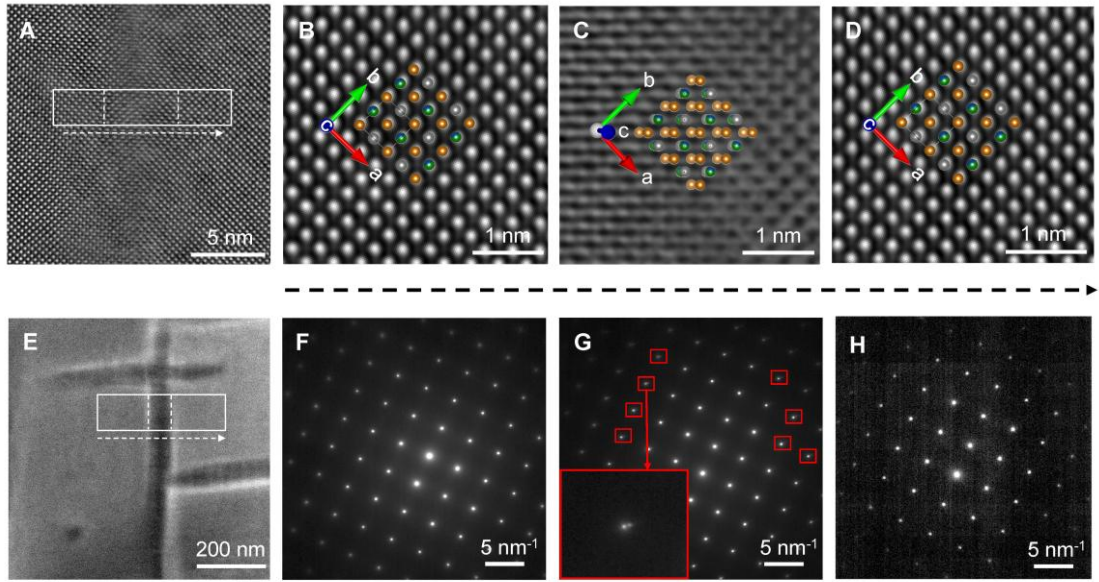

**Figure 5: Distinct crystal orientations across different regions of Mn$_{1.6}$Zn$_{0.4}$Sb. (A)** HAADF-STEM image of Mn$_{1.6}$Zn$_{0.4}$Sb, with a white rectangle highlighting an area containing both the majority regions and the stripes. **(B)-(D)** Magnified HAADF-STEM images of the left majority region, stripes, and right majority region marked in **(A)**, respectively. The simulated Mn$_{1.6}$Zn$_{0.4}$Sb atomic arrangements with dots representing atoms are overlaid on the HAADF-STEM image. (The yellow/green dots: Mn1/Mn2 atoms, blue dots: Zn atoms, and silver dots: Sb atoms). **(E)** TEM image of Mn$_{1.6}$Zn$_{0.4}$Sb, where the white rectangle highlights both the majority regions and the stripes. **(F)-(H)** SAED patterns corresponding to the left majority region, stripes, and right majority region within the rectangle in panel **(A)**, respectively.

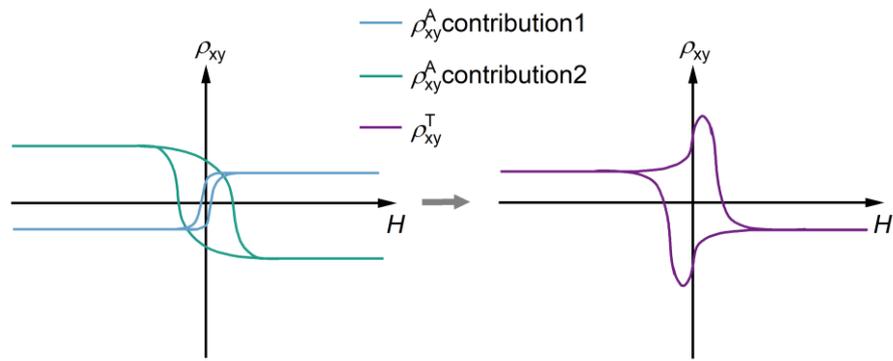

**Figure 6: Schematic illustration of the mimicry of the topological-Hall-like effect.** The topological-Hall-like signal can result from the superposition of multiple anomalous Hall loops originating from the normal region and defects with distinct magnetic easy axes or planes.

Supporting Information

# Investigating the origin of topological-Hall-like resistivity in Zn-doped $Mn_2Sb$ ferrimagnet

BoCheng Yu[1], JiaLiang Jiang[2], Jing Meng[1], XiaoYan Zhu[1,3], Jie Ma[4], HaiFeng Du[5,6,7], QingFeng Zhan[1], Jin Tang[2], Yang Xu[1], and Tian Shang[1,8]

[1]Key Laboratory of Polar Materials and Devices (MOE), School of Physics and Electronic Science, East China Normal University, Shanghai, 200241, China

[2]School of Physics and Optoelectronic Engineering Science, Anhui University, Hefei, 230601, China

[3]State Key Laboratory of Infrared Science and Technology, Shanghai Institute of Technical Physics, Chinese Academy of Sciences, Shanghai, 200083, China

[4]Key Laboratory of Artificial Structures and Quantum Control, Shenyang National Laboratory for Materials Science, School of Physics and Astronomy, Shanghai Jiao Tong University, Shanghai, 200240, China

[5]Institutes of Physical Science and Information Technology, Anhui University, Hefei, 230601, China

[6]Anhui Province Key Laboratory of Condensed Matter Physics at Extreme Conditions, High Magnetic Field Laboratory, HFIPS, Anhui, Chinese Academy of Sciences, Hefei, 230031, China

[7]Science Island Branch of Graduate School, University of Science and Technology of China, Hefei, 230026, China

[8]Chongqing Key Laboratory of Precision Optics, Chongqing Institute of East China Normal University, Chongqing, 401120, China

**Correspondence:** JiaLiang Jiang (jjl2024@ahu.edu.cn) | Yang Xu (yxu@phy.ecnu.edu.cn) | Tian Shang (tshang@phy.ecnu.edu.cn)

# Figures

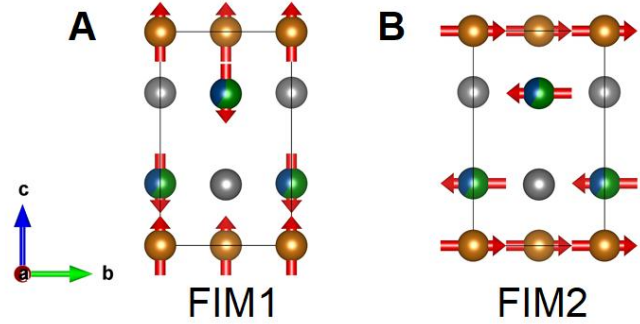

**Figure S1 Magnetic structures for the FIM1 and FIM2 phases.** The FIM1 phase **(A)** exhibits ferrimagnetic order along the c-axis, whereas the FIM2 phase **(B)** exhibits ferrimagnetic order within the ab-plane.

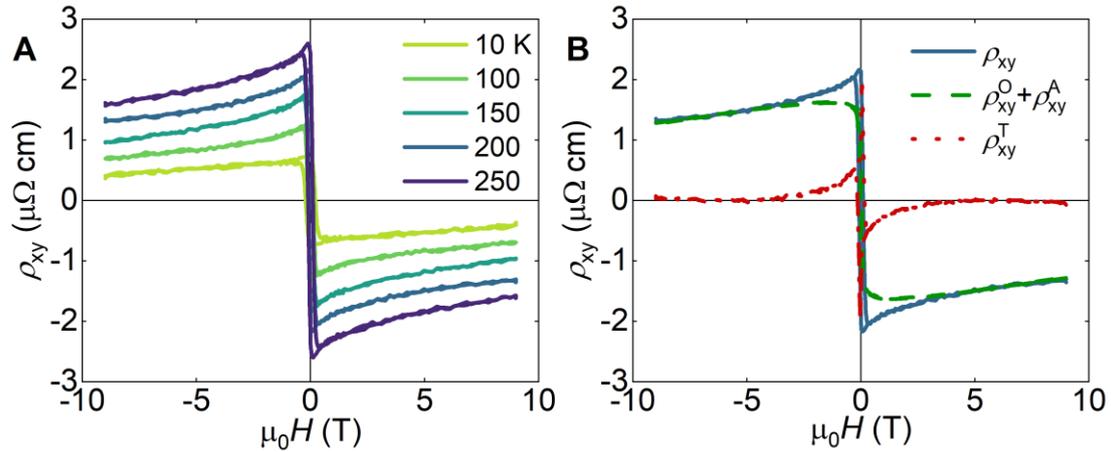

**Figure S2 Field-dependent Hall resistivity and its analysis of $Mn_{1.6}Zn_{0.4}Sb$. (A)** Field-dependent Hall resistivity $\rho_{xy}(H)$ collected at various temperatures for the $Mn_{1.6}Zn_{0.4}Sb$ single crystal in the full range of magnetic field from -9 T to 9 T. **(B)** Analysis of Hall resistivity at $T = 200$ K. Solid, dashed, and dash-dotted lines represent the measured total Hall resistivity $\rho_{xy}$, the sum of ordinary all resistivity and anomalous Hall resistivity $\rho_{xy}^O + \rho_{xy}^A$, and the topological-Hall-like resistivity $\rho_{xy}^T$, respectively.

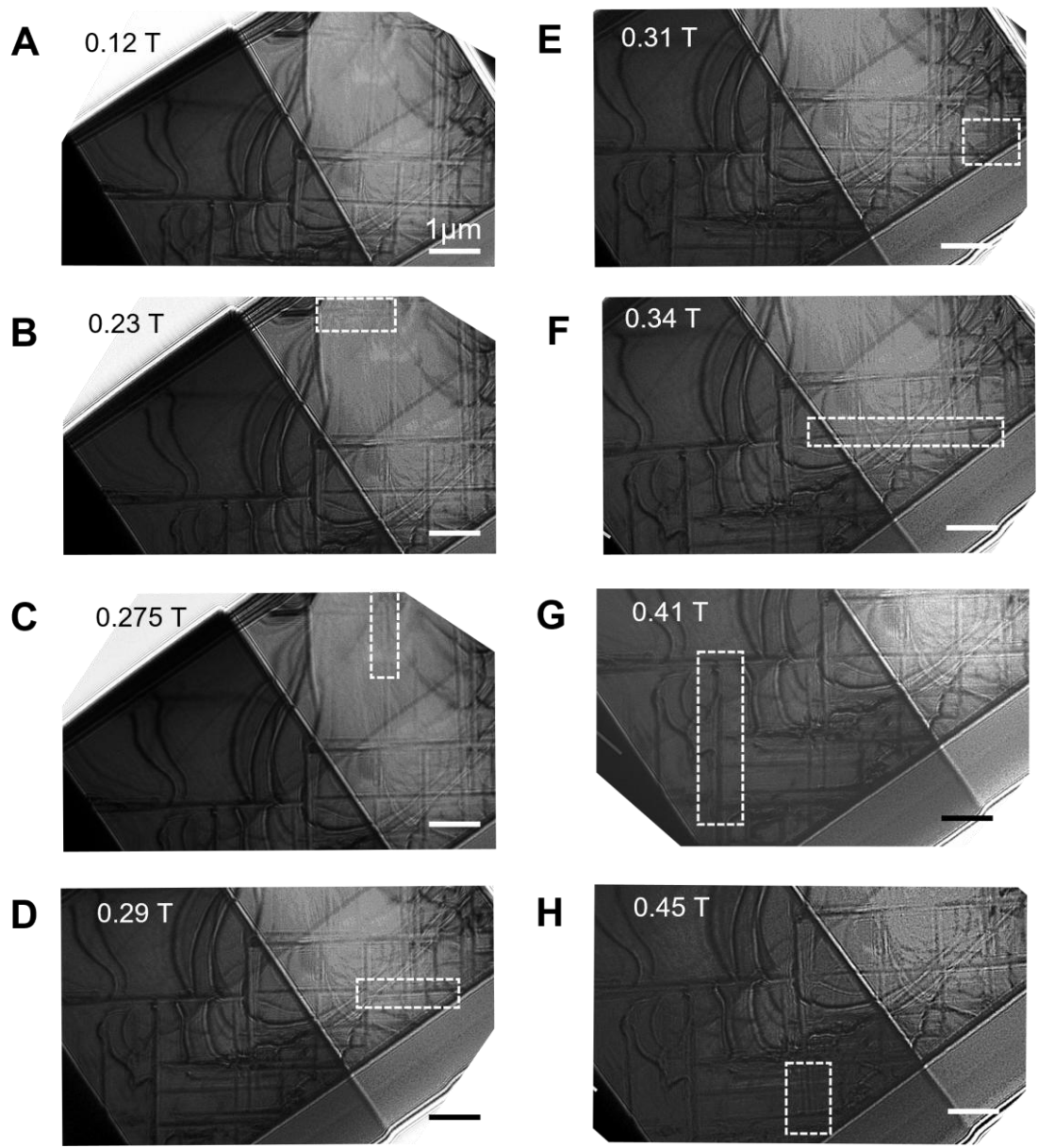

**Figure S3 LTEM images of $Mn_{1.6}Zn_{0.4}Sb$ single crystal collected at $T = 100$ K in different magnetic fields applied along the *c*-axis.** With increasing magnetic field from **(A)** to **(H)**, the image contrast of stripes at different locations undergoes reversal, as marked by the white dashed rectangles.

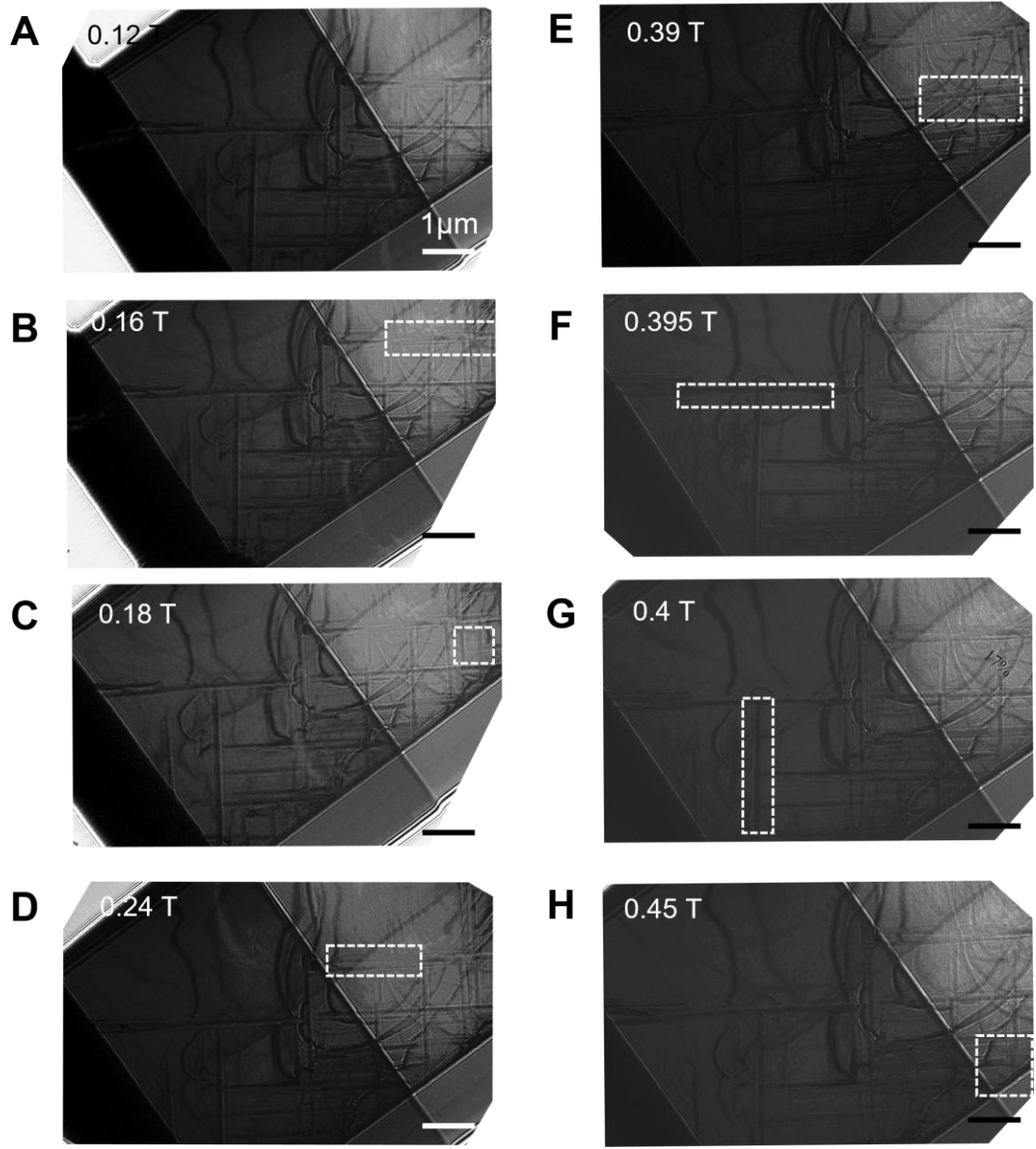

**Figure S4 LTEM images of Mn$_{1.6}$Zn$_{0.4}$Sb single crystal collected at $T = 200$ K in different magnetic fields applied along the *c*-axis.** With increasing magnetic field from **(A)** to **(H)**, the image contrast of stripes at different locations undergoes reversal, as marked by the white dashed rectangles.

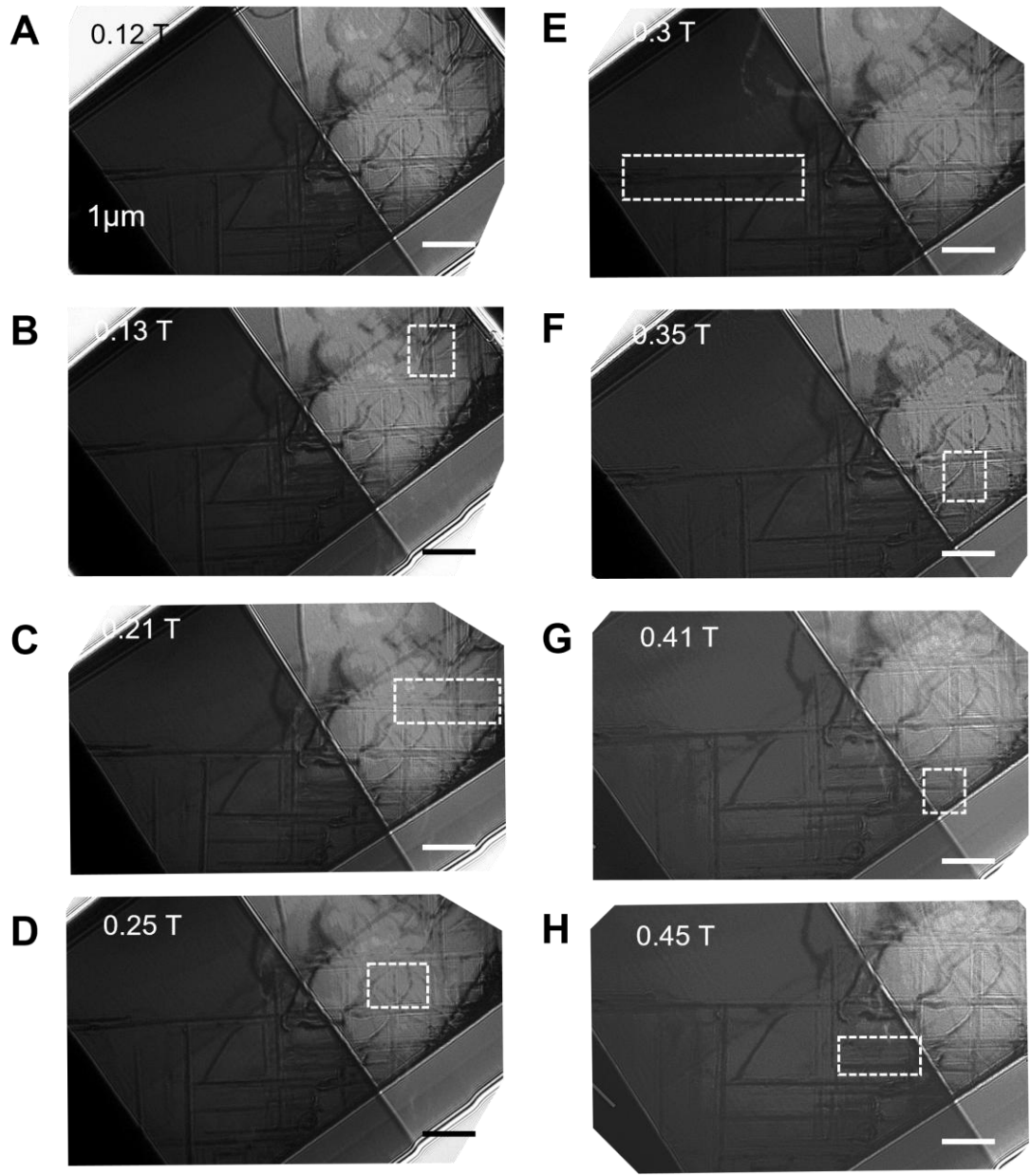

**Figure S5 LTEM images of Mn$_{1.6}$Zn$_{0.4}$Sb single crystal collected at $T = 300$ K in different magnetic fields applied along the c-axis.** With increasing magnetic field from **(A)** to **(H)**, the image contrast of stripes at different locations undergoes reversal, as marked by the white dashed rectangles.

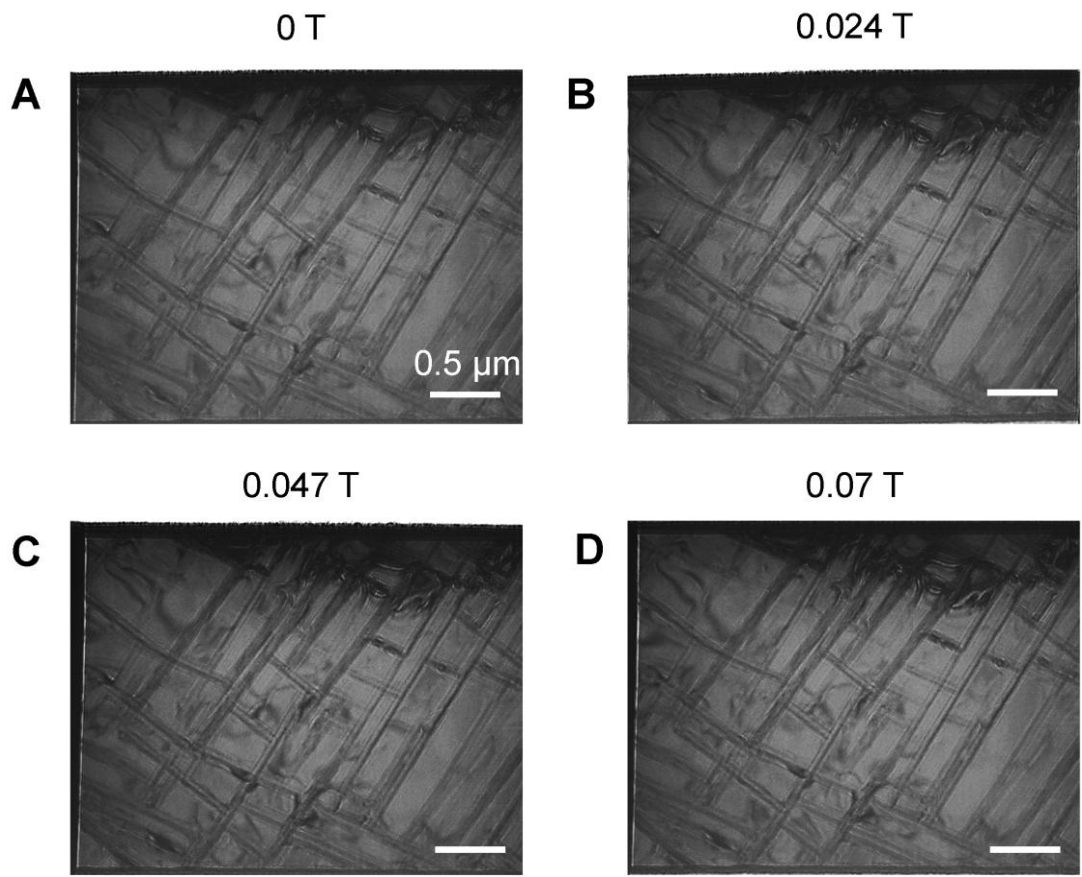

**Figure S6 LTEM images of Mn$_{1.6}$Zn$_{0.4}$Sb single crystal collected at $T = 200$ K in small magnetic fields applied along the *c*-axis. (A)–(D)** Even in the field region where the extracted Hall anomaly is most pronounced (< 0.1 T), no magnetic domain changes or chiral spin textures were observed across the entire sample area.

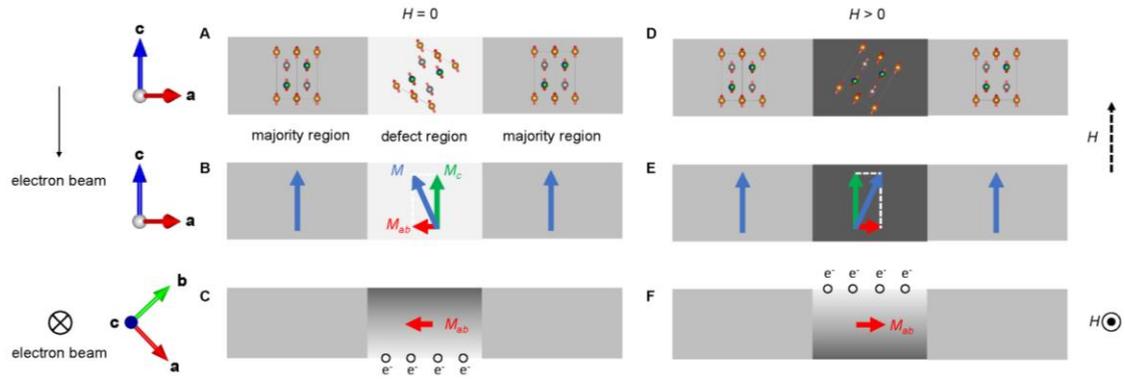

**Figure S7 Field-induced spin reorientation and LTEM contrast reversal.** Possible configurations of spin structures (**A**), net magnetic moment directions (**B**), and corresponding black-white contrast (**C**) in the majority and defect regions under zero field at $T > T_{SR}$. The analogous results after applying a magnetic field along the *c*-axis of the majority region are shown in (**D**), (**E**), and (**F**). The left coordinate axis indicates the crystal orientation of the majority region. Panels (**A**), (**B**), (**D**), and (**E**) are viewed along the (1 -1 0) direction; panels (**C**) and (**F**) are viewed along the (0 0 1) direction. The directions of the electron beam and magnetic field are also indicated.

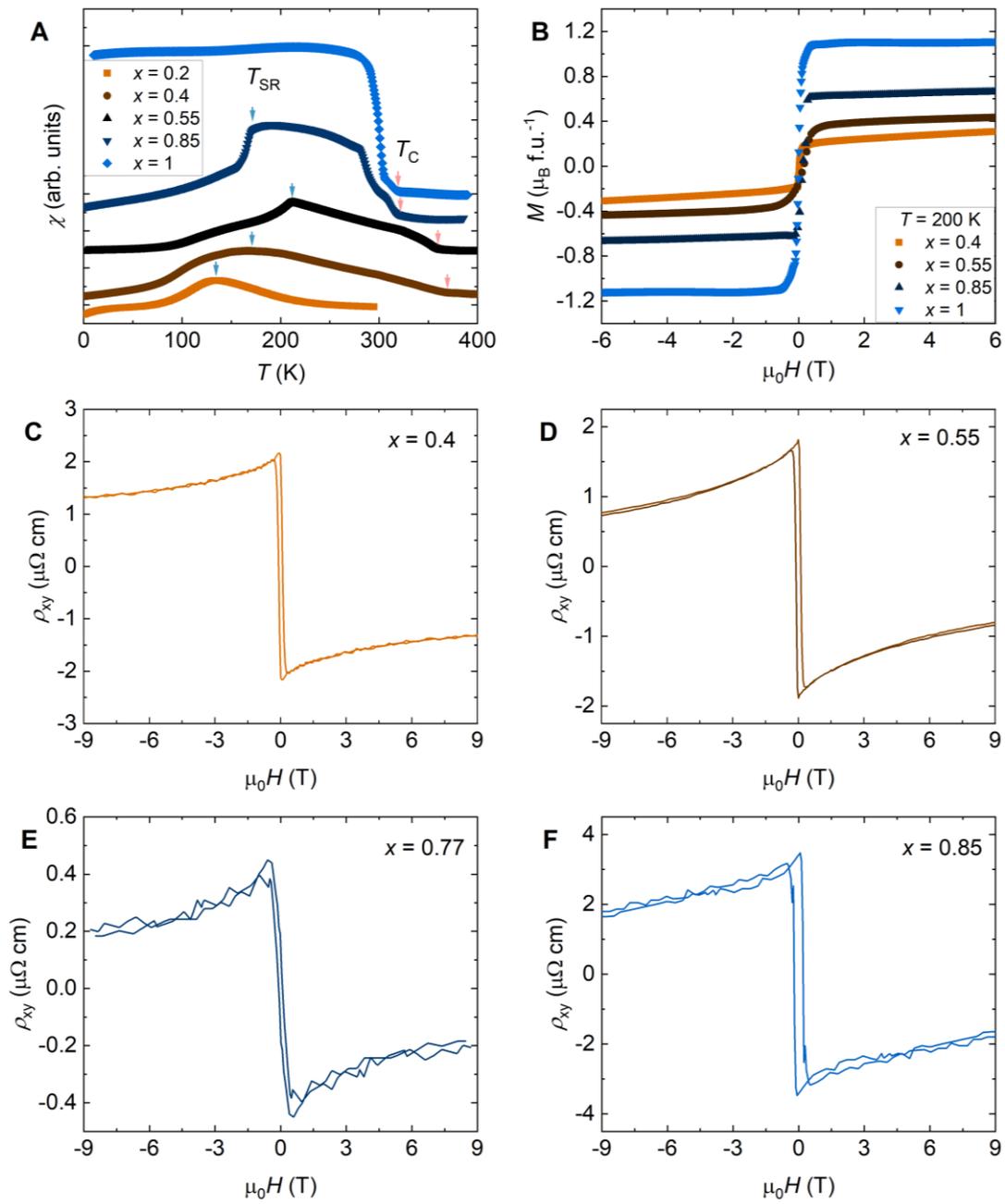

**Figure S8: Comparison of magnetic and transport properties across Zn doping levels. (A)** Temperature dependence of magnetic susceptibility for Mn$_{2-x}$Zn$_x$Sb measured with magnetic field of 0.1 T applied along the c axis. Data for each composition are shifted for clarity. The pink and blue arrows mark the FIM transition at $T_C$ and the spin reorientation transition at $T_{SR}$. **(B)** Field-dependent magnetization $M(H)$ for Mn$_{2-x}$Zn$_x$Sb collected at 200 K. **(C)-(E)** Magnetic field dependence of Hall

resistivity $\rho_{xy}$ for x = 0.4, x = 0.55, x = 0.77, and x = 0.85 samples, measured with magnetic field applied along the c axis. The magnetic and electrical data for x = 0.2 are taken from Ref. [S1] and for x = 0.77 and 0.85 from Ref. [S2].

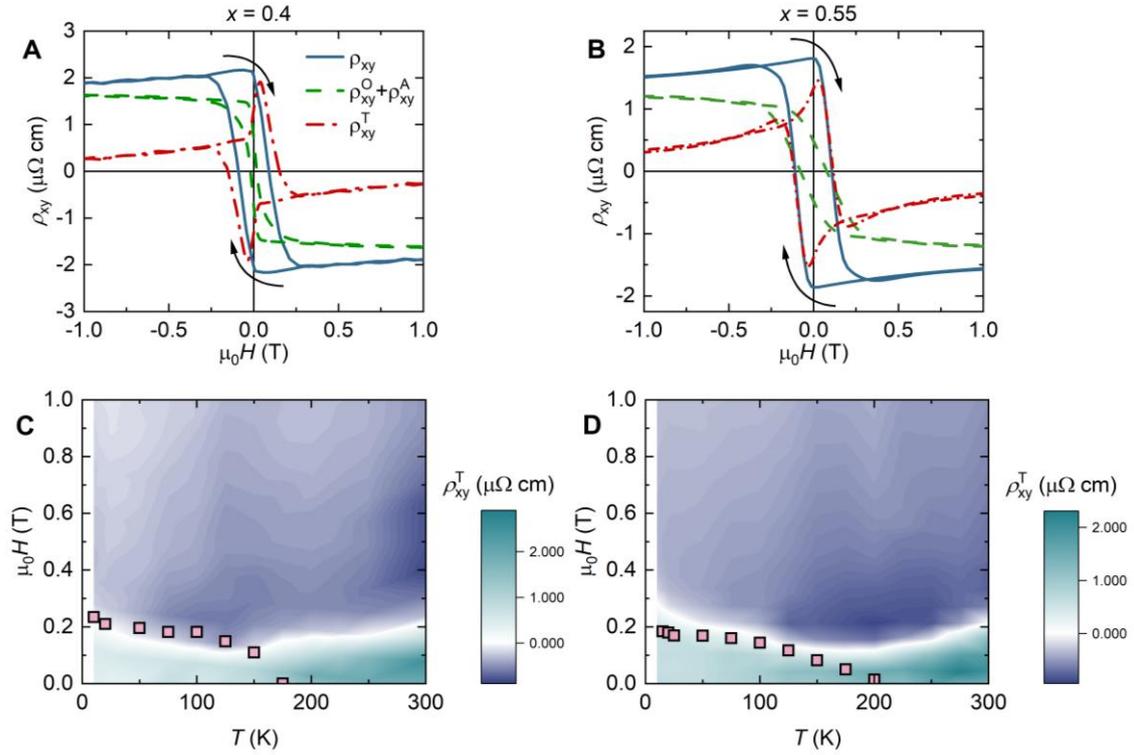

**Figure S9 Comparison of Hall resistivity analysis and magnetic phase diagrams for Mn$_{1.6}$Zn$_{0.4}$Sb and Mn$_{1.45}$Zn$_{0.55}$Sb.** Analysis of the Hall resistivity for Mn$_{1.6}$Zn$_{0.4}$Sb **(A)** and Mn$_{1.45}$Zn$_{0.55}$Sb **(B)** at $T = 200$ K. Solid, dashed, and dash-dotted lines represent the measured total Hall resistivity $\rho_{xy}$, the sum of ordinary Hall resistivity and anomalous Hall resistivity $\rho_{xy}^O + \rho_{xy}^A$, and the topological-Hall-like resistivity $\rho_{xy}^T$, respectively. Magnetic phase diagrams of Mn$_{1.6}$Zn$_{0.4}$Sb **(C)** and Mn$_{1.45}$Zn$_{0.55}$Sb **(D)** with the field applied along the *c*-axis. The critical fields (square symbols) were determined from $M(H)$ data. The back ground color in panel **(C)** and **(D)** represents the magnitude of $\rho_{xy}^T$ at various temperatures.

## Table S1

Table S1: Comparison of Hall anomaly value of $Mn_{1.6}Zn_{0.4}Sb$ and various skyrmion materials.

| Compound | Type | $\rho_{xy}^{T\,max}$ | Reference |
| --- | --- | --- | --- |
| MnSi | Flim | ~10 nΩ cm | [S3] |
| MnNiGa | Flim | ~136 nΩ cm | [S4] |
| $Mn_2PtSn$ | Flim | ~0.6 μΩ cm | [S5] |
| 4 u.c. $SrRuO_3$ | Flim | ~1.2 μΩ cm | [S6] |
| MnGe | Bulk | ~0.3 μΩ cm | [S7] |
| $Fe_3Sn_2$ | Bulk | ~0.9 μΩ cm | [S8] |
| $Gd_2PdSi_3$ | Bulk | ~3 μΩ cm | [S9] |
| **$Mn_{1.6}Zn_{0.4}Sb$** | **Bulk** | **~2.5 μΩ cm** | **This work** |

## Note S1

Taking the case of $T > T_{SR}$ as an example: in the majority region, the spins are ferrimagnetically aligned along the *c*-axis, resulting in no net in-plane magnetic moment. In the defect region, however, the distinct crystal orientation means its crystallographic *c*-axis is oriented at a random angle relative to that of the majority region [Figs. S7**A** and **D**]. Due to the presence of the defect grain boundary, magnetic moments in the defect are pinned in an arbitrary direction, producing a net in-plane magnetic moment component [Fig. S7**B**]. When a magnetic field is applied along the *c*-axis of the majority region, the majority region exhibits no contrast change as its in-plane magnetization remains zero [Fig. S7**E**]. In the defect region, however, the applied field causes the spins to revert to the crystallographic *c*-axis direction of the defect region [Fig. S7**D**]. Yet, because this direction is at a random angle relative to the *c*-axis of the majority region, a net in-plane magnetization component remains [Fig. S7**E**]. When the net in-plane moment before and after reorientation points in opposite directions—for example, in Fig. S7**C** the net in-plane magnetization in the defect region points to the left, whereas after field application in Fig. S7**F** it points to the right—a Lorentz force in the opposite direction is produced. The change in the direction of the Lorentz force causes a corresponding reversal in electron beam deflection. Bright stripes form where the electron beam converges, and dark stripes appear where it diverges. As a result of this reversal in beam deflection, the black-white contrast in the LTEM images is inverted, as shown in Figs. 4**E-H** in the main text.

**Note S2**

In our $Mn_{1.6}Zn_{0.4}Sb$ crystal, the inhomogeneity-induced Hall anomaly persists across the entire measured temperature range, reaching a maximum value of approximately 2.5 μΩ·cm around 300 K. For comparison, Table S1 summarizes the reported maximum topological Hall resistivities $\rho_{xy}^{T\,max}$ in several established skyrmion materials. The anomalous signal in our system exceeds those observed in many thin films (e.g., MnSi) and is comparable to the sizable signal reported in the bulk skyrmion material $Gd_2PdSi_3$. This demonstrates that the magnitude of the Hall signal caused by trivial structural inhomogeneity can be as large as, or even larger than, those attributed to genuine topological spin textures. Thus, the primary value of our findings is to provide a crucial cautionary note for the spintronics community. Our study demonstrates that the interpretational challenge of the THE, previously highlighted in thin films, is equally critical in bulk materials. We show that structural inhomogeneity alone can generate substantial THE-like signals, mimicking those of topological origins. This urges a re-examination of claims of chiral spin textures based solely on Hall resistivity anomalies, especially in systems with inherent microstructural disorder. Therefore, for the reliable development of spintronic devices, our work underscores the necessity of complementing transport measurements with direct microscopic verification to unambiguously distinguish between intrinsic and extrinsic origins of such Hall anomalies.

The combination of its substantial magnitude and the critical interpretational insight it provides makes the understanding of this extrinsic effect highly valuable for future research on topological magnetism.